\documentclass[12pt]{article}

\usepackage{epsfig}

\def\beq{\begin{equation}}
\def\eeq{\end{equation}}
\def\bea{\begin{eqnarray}}
\def\eea{\end{eqnarray}}

\def\gev{\, {\rm GeV}}

\def\so{s_\omega}
\def\co{c_\omega}
\def\rs1{r_{s_1}}
\def\ms1{m_{s_1}}

\def\Gsm{\Gamma^{\rm SM}}
\def\Ghid{\Gamma^{\rm hid}}
\newcommand{\gsim}{\lower.7ex\hbox{$\;\stackrel{\textstyle>}{\sim}\;$}}
\newcommand{\lsim}{\lower.7ex\hbox{$\;\stackrel{\textstyle<}{\sim}\;$}}

\begin{document}

\setlength{\baselineskip}{0.25in}


\begin{titlepage}
\noindent
\begin{flushright}
MCTP-05-91 \\
\end{flushright}
\vspace{1cm}

\begin{center}
  \begin{large}
    \begin{bf}
A Minimal Spontaneously Broken Hidden Sector and its \\
Impact on Higgs Boson Physics at the Large Hadron Collider

    \end{bf}
  \end{large}
\end{center}
\vspace{0.2cm}
\begin{center}
\begin{large}
Robert Schabinger and James D. Wells \\
\end{large}
  \vspace{0.3cm}
  \begin{it}
Michigan Center for Theoretical Physics (MCTP) \\
        ~~University of Michigan, Ann Arbor, MI 48109-1120, USA \\
  \end{it}

\end{center}

\begin{abstract}

Little experimental data bears on the question of whether there
is a spontaneously broken hidden sector that has no Standard
Model quantum numbers. Here we discuss the prospects of finding evidence for such a hidden
sector through renormalizable interactions of the Standard Model Higgs
boson with a Higgs boson of the hidden sector.  We find that the lightest
Higgs boson in this scenario has smaller rates in standard detection
channels, and it can have a sizeable invisible final state branching fraction.
Details of the hidden sector determine whether the overall width of
the lightest state is smaller or larger than the Standard Model width.  
We compute observable rates, total widths and invisible decay branching fractions
within the general framework.  We also introduce the ``A-Higgs Model'', which 
corresponds to the limit of a hidden sector Higgs boson weakly mixing with the 
Standard Model Higgs boson. This model has only one free parameter in addition to the mass
of the light Higgs state and it illustrates most of the generic phenomenology issues, 
thereby enabling it to be a good benchmark theory for collider searches.
We end by presenting an analogous supersymmetry model with similar phenomenology, which
involves hidden sector Higgs bosons interacting with MSSM Higgs bosons through
D-terms.

\end{abstract}

\vspace{1cm}

\begin{flushleft}
September 2005
\end{flushleft}

\end{titlepage}



There are many deficiencies of the Standard Model (SM), such as the hierarchy
problem, flavor problem, dark matter problem, 
cosmological constant problem, electroweak symmetry
breaking problem, CP violation problem, baryogenesis problem, etc.   Much of
the effort in particle theory has been to posit solutions to these problems and ask how they
would show up at experiments.
The presence of a hidden sector, defined here to mean extra states that have no
SM gauge charge but are charged under some other exotic gauge symmetry, does not
necessarily solve any of the problems above.  However, there are many extensions
of the SM that purport to solve these problems that do have hidden sectors
as part of their generic constructions, as is well-known to be the case for supersymmetry 
and string theory.

In this paper we take no position on what the primary motivation for a hidden
sector might be, but rather consider the most simple hidden sector possible and
ask what effect it would have on LHC phenomenology~\cite{LHC related}.
Our beginning assumptions are that there exists a hidden sector Higgs boson
$\Phi$ which can decay into undetectable hidden sector states with total width 
$\Ghid$, and that the SM Higgs state interacts only with SM
particles in the standard way.  
Our analysis will be for a condensing $\Phi$ fundamental scalar. 
This is in
contrast to other analyses which assume a singlet scalar state that has no
vacuum expectation value.  The advantage of that latter possibility is that
the hidden sector singlet state can be stable and could be the 
cold dark matter.
This interesting possibility has been studied well in the literature~\cite{singlet cdm}.
We, however, study the case of $\langle \Phi\rangle\neq 0$ since we like to think
of the hidden sector as having a rich gauge theory structure which is at least partly
broken by $\langle\Phi\rangle\neq 0$.  It is possible here that the lightest
hidden sector particle could be the dark matter, but we only require for this
analysis that the hidden sector states do not decay into SM particles within
the detector.

We comment at the start that a full precision electroweak analysis is
not possible in this study, since we are looking at generic aspects of
only Higgs physics and so are in effect summing over all the possibilities
for other hidden sector states such as $Z'$ couplings and mass that would
affect a detailed fit.  There is much freedom to have additional Higgs bosons
and $Z'$ bosons conspire to yield a reasonable fit to the precision electroweak
data~\cite{Peskin:2001rw}, and so we suggest that our approach is 
not any less applicable by
not fully considering electroweak precision data.


For our study, we will assume that the hidden sector has
at least one gauge symmetry $G_{hid}$ that is broken by a vacuum expectation
value of the Higgs boson $\Phi$.  There are two opportunities for
a renormalizable coupling between states of the hidden sector and those
of the SM.  
The first possibility for mixing between states at the renormalizable level is
kinetic mixing among the gauge bosons of $U(1)_Y$ and a $U(1)_{hid}$.  Recall
that for abelian gauge symmetry the field-strength tensor $B_{\mu\nu}=\partial_\mu B_\nu
-\partial_\nu B_\mu$ is gauge invariant, and thus an
interaction operator is allowed between the field-strengths of two different $U(1)$
symmetries,
\beq
{\cal L}_{mix}=\chi B_{\mu\nu} C^{\mu\nu}
\eeq
where $\chi$ is some dimensionless mixing parameter.
The phenomenology for theories with this kind of
interaction is interesting~\cite{Babu:1997st}; 
however, we will not focus on that here, partly because we do not
want to confine ourselves to discussions that have applicability only to
hidden sectors with abelian symmetries, and partly because the precision
electroweak fit sensitivity to this operator is higher than the one we
discuss below and being constrained as such 
would be less likely to lead to profound impacts at the LHC.

Instead, we focus on the experimental implications of the
renormalizable interaction of the SM Higgs boson with the hidden sector Higgs boson
$|H|^2|\Phi|^2$~\cite{Schabinger},
which is a 4-dimensional operator and gauge invariant. 
The Higgs boson
lagrangian under consideration for this case is
\beq
{\cal L}_{Higgs}=|D_\mu H|^2 + |D_\mu \Phi|^2 +m_H^2 |H|^2+m_\Phi^2 |\Phi|^2
-\lambda |H|^4-\rho |\Phi|^4+\eta |H|^2|\Phi|^2
\label{Lmix}
\eeq
Generically, for a stable potential that admits vevs for $H$ and $\Phi$ the
parameters $m_H^2$, $m_\Phi^2$, $\lambda$ and $\rho$ are all positive.  On the
other hand, $\eta$ is not generically required to be of one particular sign.
For simplicity, we are assuming that $\Phi$ is a
Higgs boson that breaks a $U(1)_{\rm hid}$ symmetry; however, the results that follow easily
generalize to $\Phi$ being a Higgs boson that breaks any hidden sector group 
spontaneously. 

The component fields can be written as
\beq
H=\frac{1}{\sqrt{2}}\left( \begin{array}{c} h+v +iG^0 \\ G^\pm\end{array}\right), ~~~
\Phi = \frac{1}{\sqrt{2}} (\phi+\xi +iG')
\eeq
where $v (\simeq 246\gev)$ 
and $\xi$ are vacuum expectation values about which the $H$ and
$\Phi$ fields are expanded.
The $G$ fields are Goldstone bosons absorbed by the vector bosons,
and so no physical pseudo-scalar states are left in the spectrum.  However,
the scalar spectrum has two physical states rather than just the one of
the SM.  In terms of the $\{ h,\phi\}$ interaction eigenstates, the mass
matrix one must diagonalized to obtain the two physical mass eigenstates
is
\beq
{\cal M}^2=\left( \begin{array}{cc} 2\lambda v^2 & \eta v\xi \\ 
                                   \eta v\xi & 2\rho \xi^2 \end{array}\right)
\eeq
This matrix is diagonalized by the mixing angle
\beq
\tan \omega = \frac{\eta v\xi}{(\rho \xi^2-\lambda v^2)+
\sqrt{(\rho \xi^2-\lambda v^2)^2+\eta^2 v^2 \xi^2}}
\eeq
with
\bea
h & = & \cos \omega\, s_1 + \sin\omega\, s_2 \\
\phi & = & -\sin\omega\, s_1+\cos\omega\, s_2 
\eea
The masses of the two eigenstates are then
\beq
m^2_{s_1,s_2}=(\lambda v^2+\rho \xi^2)\pm \sqrt{(\lambda v^2-\rho \xi^2)^2+\eta^2 v^2\xi^2}
\eeq
When $\eta\to 0$ we recover the two diagonal eigenstates $m_{s_1}^2=2\lambda v^2$
and $m_{s_2}^2=2\rho \xi^2$.

Within the SM, the couplings of the Higgs boson to SM particles is 
completely determined by the Higgs boson mass.  This in turn leads
to a completely specified partial width of SM Higgs decay into
any of its kinematically allowed final states.  We denote
these partial widths as $\Gsm_i(m_h)$. The total width is $\Gsm(m_h)$
with no index.  The equivalent total width of the hidden sector Higgs
boson is $\Ghid(m_\phi)$.  In other words, $\Gsm$ and $\Ghid$ are the
widths of the SM Higgs boson and the hidden sector boson, respectively, in the
case of no mixing.  We assume that all final states of $\Ghid$
are invisible to particle detectors. 

When $h$ and $\phi$ mix to form eigenstates $s_1$ and $s_2$ the decays
will in general be to both SM states and hidden sector states.  For the sake
of simplicity in writing subsequent formula, we assume that $m_{s_1}<  m_{s_2}$
and we write $\co \equiv \cos\omega$ and $\so \equiv \sin\omega$.
If $m_{s_2}>2m_{s_1}$ the decay $s_2\to s_1+s_1$ would be allowed kinematically.
The partial width of this decay depends on the parameters of the mixing
lagrangian of eq.~\ref{Lmix}.  If we isolate the $\Delta {\cal L}_{mix}=
\mu s_1^2s_2$ contribution in the lagrangian, where $\mu$ has dimensions of
mass and is determined by the vevs of the $H$ and $\Phi$ fields as well
as their dimensionless interaction coefficients, we find
\beq
\mu=\frac{\eta}{2}\left( \xi \co^3+v\so^3\right)+(3\lambda-\eta)v\co^2\so
+(3\rho-\eta)\xi\co\so^2.
\eeq
(Note, when the $\eta$ dependence of $\so$ is taken into account, $\mu\to 0$
when $\eta\to 0$, as it should.)
The resulting partial width of $s_2\to s_1s_1$ is
\beq
\Gamma(s_2\to s_1s_1)= \frac{|\mu|^2}{4\pi m_{s_2}}
\sqrt{1-\frac{4m_{s_1}^2}{m_{s_2}^2}}
\eeq
These Higgs-to-Higgs decays are tell-tale signs of Higgs mixing operator(s).

Production of $s_1$ and $s_2$ is entirely determined by the strength
of their interactions with respect to the SM.  In the case of $VV\to h$,
where $VV=gg,WW,ZZ$, we know that to first approximation it is proportional
to the Higgs boson partial width 
\beq
\sigma(VV\to h)(m_h) \propto \Gamma(h\to VV)(m_h)
\eeq
Therefore, the production cross-sections of $s_1$ is related to that
of the SM simply by $\co^2$
\bea
\sigma(VV\to s_1)(m_{s_1})=\co^2\sigma(VV\to h)(m_{s_1})
\eea
For $s_2$ the ratio is $\so^2$.
One should take note that this implies that there is no state in the
spectrum that has a production cross-section as large as the SM
Higgs boson.  

The branching fractions of the lighter state $s_1$ into a SM final state $i$
is
\bea
B_i(s_1)=\frac{\co^2\Gsm_i(m_{s_1})}{\co^2\Gsm(m_{s_1})+\so^2\Ghid(m_{s_1})}
\eea
If $\Ghid(m_{s_1})\simeq 0$, which would be the case if there were no hidden
sector states light enough for the mixed $s_1$ boson to decay into, the branching
fraction would be the same as that of the SM.  The overall width, 
on the other hand, would be
smaller by a factor of $\co^2$.
This case is equivalent to the ``universal suppression of Higgs boson
observables'' discussed in ref.~\cite{Wells:2002gq}.  In this case, light and narrow width
Higgs bosons ($115\gev \lsim m_{s_1}\lsim 160\gev$) become extraordinarily difficult
to find at colliders.  For a heavier $s_1$ boson, the smaller cross-section associated
with $\co^2$ suppression will be compensated for somewhat by the more narrow
width.  Discerning $s_1\to ZZ\to 4l$ above background is of course aided by 
a smaller width if the width is well above the detector's invariant
mass resolution.

If $\Ghid(m_{s_1})\neq 0$, the lightest Higgs state may decay predominantly
into undetectable particles depending on how large $\Ghid$ and $\so$ are.
The invisible branching fraction is 
\beq
B_{\rm inv}(s_1)=\frac{\so^2\Ghid(m_{s_1})}{\co^2\Gsm(m_{s_1})+\so^2\Ghid(m_{s_1})}
\eeq
which approaches 1 when $\so^2\Ghid\gg \co^2\Gsm$.
In this case, one has the double problem of suppressed production plus 
invisible decays.  For the most difficult case of the light Higgs boson,
the SM decay width is already quite small by accident (i.e., when $m_b\ll m_{s_1}<2m_W$),
and so it would not take much for $\Ghid$ to dominate over $\Gsm$, resulting
in an invisibly decaying Higgs boson.  The detectability of this kind of state
at a hadron collider is quite challenging even without the production suppression,
although we expect that a high-energy $e^+e^-$ collider would have relatively 
little trouble with the invisible decay aspect~\cite{linear invisible}.

All of the above discussion for $s_1$ applies to the case of $s_2$ except
the mixing factors change in an obvious way, 
and there is the new possibility of $s_2\to s_1s_1$ decays.
Thus, the production cross-section is suppressed compared to the SM by
\beq
\sigma(VV\to s_2)=\so^2\sigma(VV\to h)
\eeq
and the branching fractions into SM final states are
\beq
B_i(m_{s_2})=\frac{\so^2\Gsm_i(m_{s_2})}{\so^2\Gsm_i(m_{s_2})+\co^2\Gsm_i(m_{s_2})
+\Gamma(s_2\to s_1s_1)}
\eeq
Similarly, the branching fractions to invisible final states and $s_1s_1$
final states, which subsequently decay into a myriad of possibilities,
are determined by 
\beq
B_{\rm inv}(s_2)=\frac{\co^2\Ghid(m_{s_2})}{\co^2\Gsm(m_{s_2})+\so^2\Ghid(m_{s_2})
+\Gamma(s_2\to s_1s_1)}~~~{\rm and}
\eeq
\beq
B(s_2\to s_1s_1)=\frac{\Gamma(s_2\to s_1s_1)}{\co^2\Gsm(m_{s_2})+\so^2\Ghid(m_{s_2})
+\Gamma(s_2\to s_1s_1)}
\eeq

Let us focus on the detectability of the lightest Higgs mass eigenstate
$s_1$.  The phenomenology of this case is determined by three input
parameters:
\beq
m_{s_1},~ \so^2,~{\rm and}~ \rs1\equiv \frac{\Ghid(m_{s_1})}{\Gsm(m_{s_1})}
\eeq
In terms of these input parameters, 
the total rate of Higgs-mediated events at a hadron collider,
such as $gg\to h\to \gamma\gamma,ZZ,WW,t\bar t,\ldots$ is related to the
SM rate by
\beq
\frac{\sigma_iB_j}{\sigma^{\rm SM}_i B^{\rm SM}_j}
= \frac{(1-\so^2)^2}{1-(1-\rs1)\so^2}
\eeq
The index $i$ refers to the initial state that created the Higgs boson,
and $j$ refers to the states into which the Higgs boson decays.
The total width of the $s_1$ Higgs boson, which determines the broadness
of the reconstructed invariant mass peak, is related to the SM width by
\beq
\frac{\Gamma(m_{s_1})}{\Gsm(m_{s_1})} = 1-(1-\rs1)\so^2
\eeq
Finally, the branching fraction of the $s_1$ into hidden sector (invisible)
final states is
\beq
B_{\rm inv}(\ms1)=\frac{\so^2\rs1}{1-(1-\rs1)\so^2}
\eeq

\begin{figure}[t]
\centering
\includegraphics*[width=13cm]{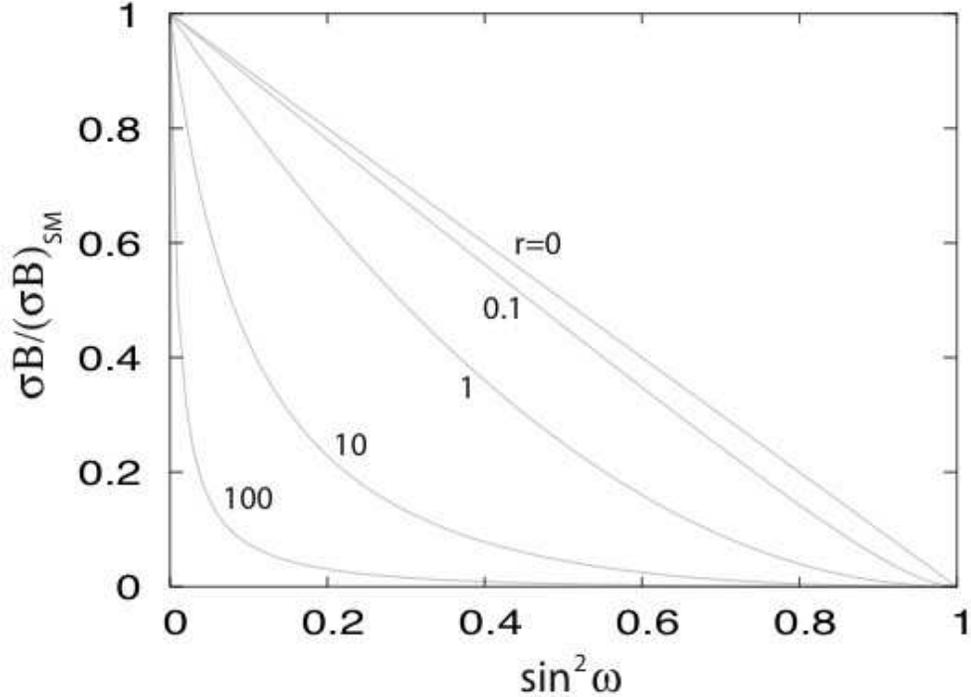}
\caption{The rate of Higgs boson observables $\sigma_iB_j$ relative
to that of the SM for various values of $r=\Ghid/\Gsm=0,0.1,1,10,100$
(from top to bottom).  
Note, the rate never exceeds that of the SM,
thus making detection of hidden sector mixed Higgs bosons more
challenging than detection of the SM Higgs boson.}
\label{Higgs rate}
\end{figure}

\begin{figure}[t]
\centering
\includegraphics*[width=13cm]{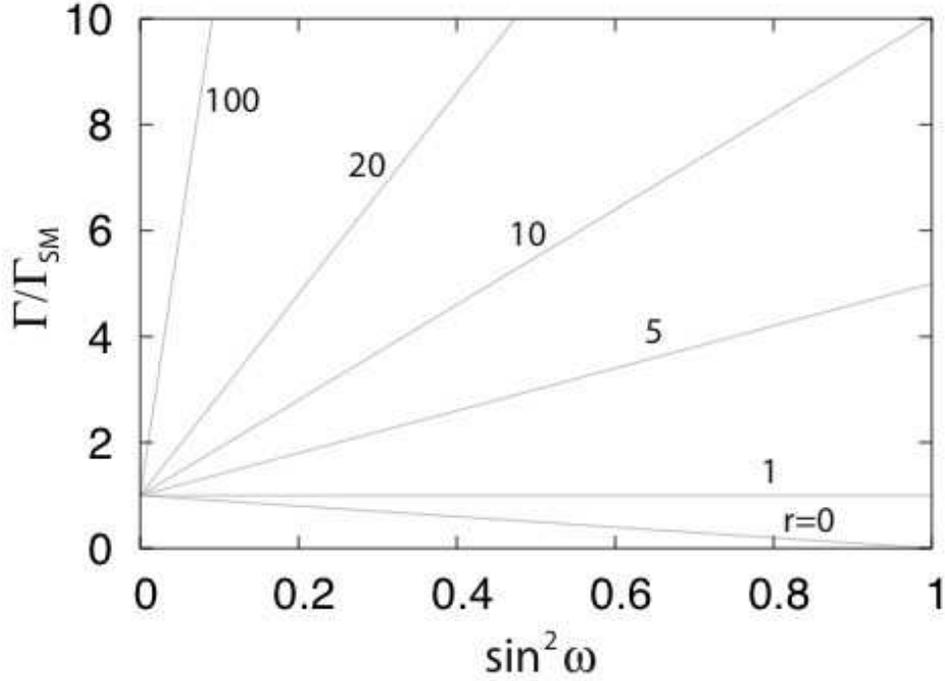}
\caption{The total width compared to that of the SM total width
for various values of $r=\Ghid/\Gsm=0,1,5,10,20, 100$ (from bottom to top). }
\label{Higgs width}
\end{figure}

\begin{figure}[t]
\centering
\includegraphics*[width=13cm]{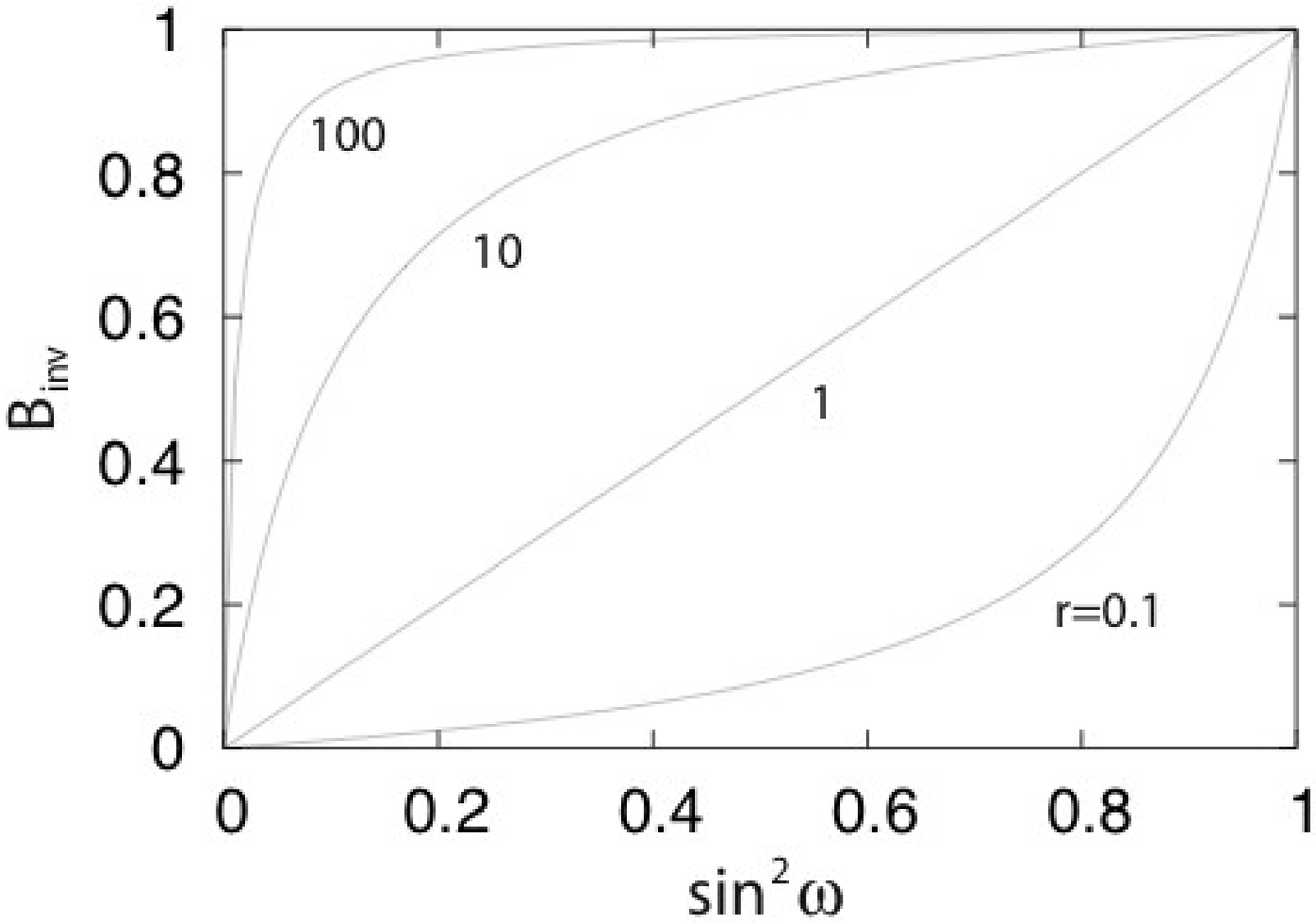}
\caption{The invisible branching fraction of the lightest Higgs boson $s_1$
for various values of $r=\Ghid/\Gsm=0.1,1,10,100$ (from bottom to top).  The
contour of $r=0$ corresponds to the $B_{\rm inv}=0$ line along the $x$-axis.}
\label{invisible width}
\end{figure}

In Figs.~\ref{Higgs rate}, \ref{Higgs width} and \ref{invisible width} 
we plot for various values of $\rs1$ the
three observables $\sigma_iB_j/(\sigma_iB_j)_{\rm SM}$, $\Gamma/\Gsm$
and $B_{\rm inv}$.  In Fig.~\ref{Higgs rate} we see that the rate for Higgs boson
induced observables never exceeds that of the SM.  This obviously
makes detection of the $s_1$ Higgs boson
in the standard channels much more difficult than detection of the
SM Higgs boson.  

In Fig.~\ref{Higgs width} we see that the total width can be quite a bit larger
than the SM Higgs boson.  Combining this fact with the results of Fig.~\ref{Higgs rate},
which showed that the rate is always suppressed, implies that 
discovering the $s_1$ in the standard detection channels at the LHC 
may take significantly more luminosity and care than is required for the 
SM Higgs boson.  On the other hand, if $r<1$, an interesting experimental
question comes into play.  In that case, as $\so^2$ increases, the observable
rate into standard channels goes down (bad for detection) whereas the 
width decreases (good for detection).  It is a technical experimental question
as to whether there is any mass range for $s_1$ where the narrowing of the
peak helps more than the dropping of the rate hurts.  We suppose that if that
ever is the case it would be in the high mass region, well above the invariant
mass resolution of the detector.

In Fig.~\ref{invisible width} we plot the invisible width of the 
$s_1$ Higgs boson.  For large values of $\so^2$ the hope of detecting
the $s_1$ Higgs boson in an invisible channel is quite small, since the
total rate is small and invisible final states are notoriously difficult
at hadron colliders~\cite{lhc 1 invisible higgs,
lhc 2 invisible higgs,tevatron invisible higgs}.  
Even for rather small values of $\so^2$ the invisible Higgs rate
could be the most important signal for the light $s_1$ boson
if $\Ghid\gg \Gsm$.  In that case, the $s_2$ boson would also
like to decay invisibly, making detection of {\rm any} Higgs boson
of the theory quite challenging.

There is an interesting limit of this framework
to analyze that is approximated by $\Ghid\gg \Gsm$ (i.e., $r\to \infty$) 
and $\so^2\to 0$, but $r\so^2\to A$. In this case,
\beq
\frac{\sigma_i}{\sigma^{\rm SM}_i}=1,~~~
\frac{\sigma_iB_j}{\sigma^{\rm SM}_i B^{\rm SM}_j}
= \frac{1}{1+A},~~~
\frac{\Gamma(m_{s_1})}{\Gsm(m_{s_1})} = 1+A,~~~{\rm and}~~~
B_{\rm inv}(\ms1)=\frac{A}{1+A}
\label{A-Higgs}
\eeq
The $s_2$ state is not produced in this limit.
Although the production cross-section for $s_1$ is the same as the SM
Higgs boson, the total rate into standard detectability channels, such
as $\gamma\gamma$ or $ZZ$, falls as $A$ increases.  At the same time
the total width increases, making these standard channels even more challenging.
On the other hand, as $A$ increases, the invisible branching fraction increases,
and the techniques for discovering invisibly decaying Higgs bosons 
become important~\cite{lhc 1 invisible higgs,
lhc 2 invisible higgs,tevatron invisible higgs}.

We consider
this model to be interesting physically, as it implies that there is
very little interaction between our sector and the hidden sector but
the hidden sector has large decay width into its own hidden sector
states compared to the SM Higgs boson. This is especially interesting
for a rather light Higgs boson ($\lsim 160\gev$) which as we discussed
above accidentally has a small width into SM states.  
Thus, the ``A-Higgs model'' of eq.~\ref{A-Higgs}
would be interesting to study in detail in our view since
it is motivated physically, has only two free parameters ($\ms1$ and $A$) and
illustrates much of the generic phenomenology relevant to hidden sector Higgs mixing
(reduced rates, increased widths and invisible decays).

Finally, we comment on the supersymmetric analogue of this case.  
By the definition of hidden sector we have given above, a prime
supersymmetry candidate would be the $S$ field in $\Delta W=\lambda SH_uH_d$.
If $S$ is charged under another $U(1)$ then its vacuum expectation value
would break that symmetry, which fits into our general discussion well.
An added bonus is that the vev would generate the $\mu$ term through
$\lambda\langle S\rangle=\mu$.  This scenario has been studied quite
extensively in the literature~\cite{singlet susy}, and we do not have anything additional
to say regarding its LHC phenomenology.

A slightly ``more hidden'' sector in supersymmetry, which is closer to the
spirit of the SM hidden sector discussed above, is a field $\Phi$ that
is charged under a new $U(1)_{\rm hid}$ and participates in a $D$-term interaction
with the SM Higgs bosons, but has no gauge-invariant, renormalizable interaction
in the superpotential.  Of course, a $D$-term interaction would require the $H_u$ and $H_d$
fields to be charged under $U(1)_{\rm hid}$ as well.  Although $\Phi H_uH_d$ is not
allowed in the superpotential by assumption here, the $D$ term interaction
between the states yields a mixing in close analogy with our SM case above:
\beq
V=\frac{g_{\rm hid}^2}{2}\left( Q_{u}|H_u|^2+Q_{d}|H_d|^2+|\Phi|^2 +\cdots \right)^2.
\eeq
For simplicity we are normalizing the $U(1)_{\rm hid}$ such that $\Phi$ has charge $+1$.
Assuming the MSSM soft lagrangian for $H_u$ and $H_d$, and adding a soft mass
for $\Phi$ such that $\langle \Phi\rangle\neq 0$, we can construct the CP-even
tree-level Higgs mass matrix in the $\left\{ h_u,h_d,\phi\right\}$ basis, where
${\rm Re}(H_u)=(h_u+v)/\sqrt{2}$, ${\rm Re}(H_d)=(h_d+v_d)/\sqrt{2}$ and
${\rm Re}(\Phi)=(\phi+\xi)/\sqrt{2}$:
\beq
{\cal M}^2=\left( 
\begin{array}{ccc}
m^2_Ac_\beta^2+(m^2_Z+m^2_{Z'}\gamma^2Q_{u}^2)s_\beta^2 &
-\left(m^2_A+m^2_Z-m^2_{Z'}\gamma^2 Q_{u}Q_{d}\right)s_\beta c_\beta &
m^2_{Z'}\gamma Q_{u}s_\beta \\
-\left(m^2_A+m^2_Z-m^2_{Z'}\gamma^2Q_{u}Q_{d}\right)s_\beta c_\beta &
m^2_A s_\beta^2+(m^2_Z+m^2_{Z'}\gamma^2Q_{d}^2)c_\beta^2 &
m^2_{Z'}\gamma Q_{d}c_\beta \\
m^2_{Z'}\gamma Q_{u}s_\beta & m^2_{Z'}\gamma Q_{d}c_\beta & m^2_{Z'}
\end{array}\right)
\eeq
In terms of the gauge couplings and vevs, 
$m^2_{Z'}=g_{\rm hid}^2\xi^2$ and $\gamma^2=(v^2_u+v^2_d)/\xi^2$.  The values of
$c_\beta$ and $s_\beta$ are obtained from $t_\beta=\tan\beta=v_u/v_d$.
This matrix is now $3\times 3$ in contrast to the $2\times 2$ matrix of the MSSM
due to the additional Higgs field.

The upper bound of the lightest Higgs boson state of this supersymmetric
theory can be obtained by finding the lightest eigenvalue of the upper two-by-two
matrix, which is
\beq
m_{s_1}^2\leq m^2_Zc_\beta^2+m_{Z'}^2\gamma^2(Q_dc^2_\beta+Q_u s^2_\beta)^2+\Delta_{\rm rad}
\eeq
where $\Delta_{\rm rad}$ is the quantum correction whose leading contribution comes
from the top-stop loops of the MSSM.  
For heavy $m_A$ and heavy $m_{Z'}$ the bound given above becomes
the actual eigenvalue.
In the MSSM, $\Delta_{\rm rad}$ generally 
needs to be above $\sim (70\gev)^2$ in order for $m_h$
to be above the experimental limit of $114\gev$.  Such a high radiative
correction is not easy to obtain within weak-scale supersymmetry, and thus
strains our ordinary view of naturalness~\cite{susy higgs}.
The lightest Higgs mass in the MSSM has a maximum mass
of about $130\gev$ for TeV-scale supersymmmetry.  Furthermore, much of the supersymmetry
parameter space that is considered ``natural'', which is associated with lighter
weak-scale superpartner masses,
leads to a prediction of the mass lighter than the current
experimental limit of $114\gev$.  Thus, the extra $m_{Z'}^2\gamma^2$
contribution to the mass eigenvalue is welcome. Nevertheless, we still expect the 
lightest Higgs mass to stay relatively light even in this theory since
$m_{Z'}^2\gamma^2\sim m_Z^2$.  Therefore, the Higgs boson width into SM states
remains narrow (again, $m_b\ll m_{s_1}<2m_W$), and the
Higgs boson detection phenomenology of this supersymmetric theory
is very similar to that of the SM-variant that we discussed earlier.
All equations that were derived above for the SM hidden sector case 
carry forward with little change except to take care of the extra mixing
angles in a $3\times 3$ matrix.  

The simple illustrative
case of large $\Ghid$ and small mixing within this supersymmetric
context yields a similar
result to what we found in the non-supersymmetric case:
the lightest Higgs boson will have large suppressions of standard detection final
state rates at the LHC, and there will be a premium on good analyses
that keep free the total rates, the widths of the Higgs bosons and the
branching fraction into invisible final states.
The ``A-Higgs model'' discussed earlier is the simplest illustration of
all these effects.


\section*{Acknowledgements}
This work was supported in part by the Department of Energy, the Michigan
Center for Theoretical Physics, and the
NSF REU program.  We thank J. Kumar, S. Martin and M. Toharia for helpful discussions.

\end{document}